\documentclass[epj]{svjour}
\usepackage{graphicx} 

\begin{document}

\title{Quantum Monte Carlo Calculation for the neutron-rich Ca isotopes}

\author{S. Gandolfi\inst{1,2} \and F. Pederiva\inst{2,3,4} \and S. a Beccara\inst{2,3}}

\institute{
S.I.S.S.A., via Beirut 2-4, 34014 Trieste, (Italy)
\and
INFN, Gruppo Collegato di Trento, via Sommarive 14, 38050 Povo, Trento (Italy)
\and
Dipartimento di Fisica dell'Universit\'{a} di Trento, via Sommarive 14, 38050 Povo, Trento (Italy)
\and
INFM {\sl DEMOCRITOS} National Simulation Center, via Beirut 2-4, 34014 Trieste (Italy)
}

\abstract{
We computed ground-state energies of calcium isotopes from $^{42}$Ca to $^{48}$Ca 
by means of the Auxiliary Field Diffusion Monte Carlo (AFDMC) method. 
Calculations were performed by replacing the $^{40}$Ca core 
with a mean-field self consistent potential computed using Skyrme interaction.
The energy of the external neutrons is calculated by projecting the ground-state from 
a wave function built with the single particle orbitals computed in the self consistent 
external potential. The shells considered were the $1F_{7/2}$ and the $1F_{5/2}$.
The Hamiltonian employed is semi-realistic and includes tensor, spin--orbit and 
three--body forces. 
While absolute binding energies are too deep if compared with 
experimental data, the differences between the energies for nearly all isotopes are 
in very good agreement with the experimental data.
}

\PACS{
{21.10.-k}{Properties of nuclei; nuclear energy levels} \and
{21.10.Dr}{Binding energies and masses} \and
{24.10.Cn}{Many-body theory} \and
{21.30.Fe}{Forces in hadronic systems and effective interactions}
}

\maketitle

\section{Introduction}
The study of medium-heavy nuclei using a realistic Hamiltonian is a very hard problem to attack because of the 
complexity of the nucleon-nucleon (NN) potentials and of the three-body forces (TNI). 
The calculation of properties of a nuclear systems generally starts by considering an effective 
Hamiltonian with a NN and TNI effective interactions describing the one--, two-- and three-pion exchange 
between nucleons. Typically other effects are also included in the NN interaction, that is fitted to the experimental 
scattering data.

Efficient methods have been developed to accurately solve the Schr\"odinger equation for 
few-nucleon bound states with a realistic NN interaction\cite{kamada01,viviani07}
and with a three-body force\cite{gazit06,viviani05}. However, nuclei which can be treated by these methods have very 
small mass numbers A\cite{barnea00,barnea01}.

Projection Monte Carlo methods are the best candidate to perform calculations with such potentials. 
In fact all the results obtained so far are 
in close agreement with experimental data. In turn, these procedures allowed for the development of
more and more accurate model interactions 
that are now good enough to describe light nuclei\cite{pieper02} very accurately.

All nucleons Green's Function Monte Carlo (GFMC) calculations are presently limited to systems 
with only 12\cite{pieper05} nucleons or 14\cite{carlson03} neutrons.
Medium-heavy nuclei are usually investigated with approximate many-body techniques
like Hartree-Fock\cite{xinhua97,vautherin72}, Correlated Basis Functions\cite{bisconti07}, 
Variational Monte Carlo\cite{pieper92}, or other techniques\cite{brown98,mcgrory70} 
which are less demanding from the computational point of view, but which contain uncontrolled approximations.   

The study of neutron-rich isotopes is an interesting field of investigation because their 
structure is particularly relevant to constrain properties of the crust in neutron stars\cite{gupta07}.
The impossibility to use accurate methods such the GFMC to study nuclei with intermediate mass values, in 
addition to the complexity of NN and TNI nuclear forces imposes the need to introduce a
model simpler than the full microscopic description of a nucleus including all degrees of freedom.
The study of heavy isotopes of oxygen has already been performed within a scheme in which 
the closed-shell nucleus is substituted by an external potential\cite{gandolfi06}.  
In this paper we propose a similar analysis of heavy calcium isotopes in the $1F_{7/2}$ and 
$1F_{5/2}$ shells based on the Auxiliary Field Diffusion Monte Carlo (AFDMC) method, that was
already used in several works on pure neutron systems giving results of a quality 
comparable with those of GFMC\cite{pederiva04}. 
This case presents some 
additional difficulties with respect to that of the oxygen isotopes. In fact the density 
of external neutrons is closer to that of $^{40}$Ca, and therefore core polarization effects may 
become more important. The external neutrons, treated explicitly, interact with a realistic 
two and three body potential (in particular the Argonne AV8'\cite{wiringa02} plus 
the Urbana IX\cite{pieper01}).
Recently, the AFDMC was extended to proton-neutron systems such nuclei\cite{gan07b} and nuclear 
matter\cite{gandolfi07}, but using a Hamiltonian limited to include only the main contribution coming from 
the one-pion exchange.

The aim of this paper is to verify whether the single-particle picture is still a good starting point for 
this kind of calculations and whether the AFDMC algorithm is able to account for the relevant quantum many-body correlations 
for neutron systems. In fact, for nuclei containing a number of nucleons above the closed 
shell numbers of oxygen the effect of magic numbers might vanish, and the accuracy of the shell model 
might be reduced. 
In addition, in this case, neutrons have a large angular momentum 
($l=3$) that might play an important role in the spin-orbit interaction.
It might be hard to correctly evaluate the contribution of the spin-orbit part to the total binding 
energy.
In addition, if by replacing the core of $^{40}$Ca with an external well, we should find that the separation 
energies between different isotopes were in a good agreement with experimental data, then the 
contribution to energies separation would come almost completely from NN and TNI used, and core polarization
effects can be neglected.

One could in principle include other shells having similar single particle energy, like 
the $2P_{3/2}$, in order to start from a more accurate 
trial wave function to project out the ground-state properties of the system. We will show that 
results obtained limiting the space to the $1F$ shell are already very satisfactory. The inclusion 
of shells with principal quantum number different from 1 would introduce many additional technical 
difficulties\cite{gandolfi06}.

\section{Hamiltonian and methods}
The ground-state energies of the calcium isotopes are computed starting from a non-relativistic 
Hamiltonian of the form:
\begin{eqnarray}
\hat{H} &=& T+V_1+V_2+V_3
\nonumber \\ 
&=&-\sum_i\frac{\hbar}{2m}\nabla^2_i+\sum_i V_{ext}({\vec r}_i)+
\sum_{i<j} v_{ij} + \sum_{i<j<k}V_{ijk} \,, \nonumber \\
\end{eqnarray}
where the particle index $i$ and $j$ labels only the N external neutrons. In our calculations we considered 
systems with N=2 to N=8.
The one-body potential $V_{ext}$ represents the $^{40}$Ca core; it has been obtain from Hartree-Fock 
calculations using Skyrme forces. In fact the only input of the theory 
is the set of Skyrme parameters. Based on our previous experience with oxygen isotopes we chose the set 
of parameters corresponding to the Skyrme I interaction\cite{vautherin72}. However we expect that the choice of the 
self-consistent potential only influences the absolute total binding energy of each isotope. The 
energy difference between isotopes mostly depends on the choice of the NN and TNI interactions including
correlations between the external neutrons. Our 
approximation corresponds to neglecting the effects of the interaction between external 
neutrons and the core. 

The two body interaction used belongs to the Urbana-Argonne family\cite{wiringa95}; 
it is written as a sum of operators:
\begin{equation}
v_l(i,j)=\sum_{i<j} \sum_{p=1}^l v_p(r_{ij})O^{(p)}(i,j) \,.
\label{pot}
\end{equation}
In this work we truncated the sum to include only the first 8 operators with the parameters of the 
Argonne AV8' potential\cite{argonnev18}. This contains the usual operators:
\begin{equation}
O^{p=1,8}(i,j)=(1,\vec\sigma_i\cdot\vec\sigma_j,S_{ij},
{\vec L}_{ij}\cdot{\vec S}_{ij})\times(1,\vec\tau_i\cdot\vec\tau_j) \,,
\end {equation}
where the operator $S_{ij}=3\vec{\sigma}_i\cdot\hat{r}_{ij}
\vec{\sigma}_j\cdot\hat{r}_{ij}-\vec{\sigma}_i\cdot\vec{\sigma}_j$ is the
tensor operator
and $\vec L_{ij}=-\imath\hbar\vec r_{ij}\times (\vec\nabla_i-\vec\nabla_j)/2$
and $\vec S_{ij}=\hbar(\vec\sigma_i+\vec\sigma_j)/2$ are the relative
angular momentum and the total spin for the pair $ij$.
For neutrons $\vec\tau_i\cdot\vec\tau_j=1$, and we are left with an isoscalar
potential.

The AV8' potential used in this work is a simplified version of the more accurate Argonne AV18.
It reproduces very well the binding energy of nucleons for densities smaller then or equal 
to the equilibrium one. It has been used in calculations of light nuclei
\cite{pieper01,pudliner97}, symmetric nuclear matter\cite{gandolfi07,akmal97}, neutron 
matter\cite{sarsa03}, spin polarized neutron matter\cite{fantoni01}
and neutron rich nuclei\cite{gandolfi06}. The Hamiltonian contains also the Urbana IX (UIX) 
potential\cite{pieper01}.

Calculations were performed using the AFDMC method \cite{schmidt99}.
This algorithm projects out the lowest-energy ground-state from a trial wave function $\psi_T$ by a 
propagation in imaginary time $\tau$:
\begin{equation}
\psi(\tau)=e^{-(H-E_T)\tau}\psi_T \,,
\end{equation}
and for sufficiently large $\tau$ 
\begin{equation}
\phi_0=\lim_{\tau\rightarrow\infty}\psi(\tau) \,,
\end{equation}
where $\phi_0$ is the lowest-energy component of $\psi_T$ not orthogonal to $\psi_T$.
The evolution in imaginary-time is reached by solving the integral equation including importance sampling
\begin{equation}
\label{eq:int}
\psi_T(\vec R)\psi(\vec R,\tau)=
\int d\vec R' G(R,R',\tau)\frac{\psi_T(\vec R)}{\psi_T(\vec R')}
\psi_T(\vec R')\psi(\vec R',0) \,,
\end{equation}
where $G(R,R',\tau)$ is the approximate Green's function of the system, that in the limit 
of small time-step is 
\begin{equation}
G(\vec R,\vec R',\Delta\tau)=
\left({m\over2\pi\hbar^2\Delta\tau}\right)^{3A\over2}
e^{-{m(\vec R-\vec R')^2\over2\hbar^2\Delta\tau}}
e^{-{V(\vec R)+V(\vec R')\over2}\Delta\tau} \,.
\end{equation}
Then the integral of Eq. \ref{eq:int} must be iteratively solved until the convergence is reached.

The AFDMC algorithm implements the 
usual diffusion process for the particle positions and samples the spin states of neutrons 
with a propagator written in terms of a Hubbard-Stratonovich transformation. 
This is done in order to reduce the quadratic spin dependence of nuclear 
Hamiltonians on the spin operators to an integral over auxiliary field variables, therefore 
averaging the sum over spin states. 

As in standard DMC, 
AFDMC suffers of the fermion sign problem due to the antisymmetric character of the wave function; 
in our case the importance function is complex, and we constrain walkers to propagate within regions
where the real part of the importance function has the same sign.
A detailed description of the method and of the constrained path\cite{zhang03}, used to control the fermion 
sign problem, can be found in\cite{gandolfi07c}.

\subsection{Wave function}
The wave function used both as importance and projection function for the AFDMC algorithm has 
the following form:
\begin{eqnarray} 
\psi_I(\vec R,S) =  F_J(\vec R)\ D(\vec R,S) \,,
\end{eqnarray}
where $ \vec R\equiv (\vec r_1,\dots,\vec r_N) $ and $S\equiv (s_1,\dots ,s_N)$.
The spin assignments $s_i$ consist in giving the spinor components, namely
\[ 
s_i \equiv \left(\begin{array}{c} 
u_i \\ d_i
\end{array}\right)=u_i |\uparrow\rangle + d_i |\downarrow\rangle \,, 
\]
where $u_i$ and $d_i$ are complex numbers.

The Jastrow correlation operator is given by
\begin{eqnarray}
F_J(\vec R)&=&\prod_{i<j}f_J(r_{ij}) \,,
\end{eqnarray}
while the antisymmetric part of the trial wave function is
\begin{eqnarray} 
D(\vec R,S) &=&A\left[ \phi_{\alpha_j}(\vec r_i,s_i)\right]
\end{eqnarray}
that is the Slater determinant of one--body spin--space orbitals:
\begin{eqnarray}
\phi_{\alpha}(\vec r,\sigma)=R_{n,j}(r)Y_{l,m_l}(\theta,\phi)\xi_{s,m_s}(\sigma) \,,
\end{eqnarray}
and $\alpha=\{n,j,m_j\}$.

The radial components $R_{n,j}(r)$ were obtained solving the Hartree-Fock (HF) problem with the Skyrme I
force\cite{vautherin72}. The resulting self-consistent single particle potential is used in substitution of the 
closed $^{40}$Ca core. The yielded radial functions are written in the $j,m_j$ 
base.

The angular components were chosen to be eigenfunctions of the total angular momentum operator $J$, in order 
to enforce the proper angular nodal structure and to reduce the computational load. They were obtained by applying 
projection operators of the required momentum on single-determinant wave functions, in the $l$--$s$ representation. 
Clebsch-Gordan coefficients were obtained automatically by projecting from single-particle wave function and constructing the transition matrix.

In  $^n$Ca, n=42...48 isotopes, we assume that neutrons fill only the orbitals in the
$1F_{7/2}$ and $1F_{5/2}$ shell in order to build the ground-state of correct symmetry.
The many-body states are obtained by coupling the single-particle angular momentum by
constructing eigenstates of total angular momentum $J=j_1+...+j_N$ with
N=2...8; for the states with an even number of neutrons, the ground-state
has $J=0$, while for odd neutron numbers, the ground-state has
total angular momentum $J=7/2$.
These states are in general written in terms of a linear combination of Slater determinants,
whose coefficients are determined by the symmetry of the state, and obtained by projection. Each
determinant is evaluated at the current values of the positions and spin assignments
of the nucleons in the walker $|R,S\rangle$.  

A sufficiently good representation of the ground-state of $^{46}$Ca can be obtained by building 
a two-hole state which is complementary to $^{42}$Ca.
For the $^{43}$Ca and $^{45}$Ca completely different trial wave function are needed; in 
fact the wave function for the first nucleus contains 9 determinants and the second 35. 
In principle we could use the $^{43}$Ca wave function as a three-hole state describing 
$^{45}$Ca. However, AFDMC gives in this case an energy a few MeV too high. This is probably due 
to the approximation used to deal with the sign problem. The two wave functions are 
degenerate in the eigenspace of $J^2$ but they give different AFDMC energies, because 
of the different nodal surface.

The Jastrow function $f_J$ has been taken as the scalar component of the 
Fermi Hypernetted Chain in the Single Operator Chain approximation
(FHNC/SOC) correlation operator $\hat F_{ij}$ which minimizes the energy per
particle of neutron matter at density $\rho$=0.16 fm$^{-3}$\cite{wiringa88}.
The Jastrow part of the function in our case has the only role of reducing the overlap 
of nucleons, therefore reducing the energy variance.
Since it does not change the phase of the wave function, it
does not influence the computed energy value in projections methods.
For this reason the Jastrow function has not been further
optimized for our calculations.

In order to have an accurate estimate of the ground-state energies
of the isotopes we performed several sets of runs for different values
of the imaginary time step and walker populations.
The reported values are all extrapolated to $\Delta\tau \rightarrow 0$.
Also the dependence of the result from the number of walkers used was investigated. 
Some calculations were repeated for 500 and 1000 walkers. For 
relatively long time-steps, the energy has rather large  fluctuations when
a smaller number of walkers is employed. On the other hand the average energy does
not show a clear trend outside the statistical fluctuations.
Therefore, we present all results obtained with 1000 walkers.

\section{Results}
\begin{table}[h]
\begin{center}
\begin{tabular}{||l|cc||}
\hline
isotope     & $E_{AFDMC}$ & $E_{exp}$ \\
\hline
\hline
E($^{42}$Ca) & -25.25(7) & -19.85 \\ 
E($^{43}$Ca) & -33.9(4)  & -27.78 \\
E($^{44}$Ca) & -46.3(1)  & -38.91 \\
E($^{45}$Ca) & -52.6(4)  & -46.32 \\ 
E($^{46}$Ca) & -62.9(3)  & -56.72 \\
E($^{47}$Ca) & -70.2(7)  & -63.99 \\
E($^{48}$Ca) & -80.3(8)  & -73.09 \\
\hline
\end{tabular}
\end{center}
\caption{Ground-state energy computed with AFDMC. All the energies are 
given in MeV. Experimental values are referred to the ground-state energy of 
$^{40}$Ca, taken from Ref. \cite{exp00}.}
\label{table:ris}
\end{table} 

In table \ref{table:ris} we report the AFDMC energies obtained for the isotopes
series $^{42}$Ca--$^{48}$Ca, compared with the available experimental values\cite{exp00}.

As expected, the absolute binding energies are quite different from the experimental 
results, although the
relative discrepancy never exceeds 30\%. This is a drawback of using
of an external potential for including the effects of the filled core of the
nucleus. The total binding energies are all overestimated.
This reflects the absence of a correct description of the density of neutrons at the
center of the drop, which is underestimated due to the absence of the $S$ states, giving rise to an effective
potential which is too deep for small distances of the neutrons from the center. Moreover we completely
neglect core--polarization effects.

\begin{table}[h]
\begin{center}
\begin{tabular}{||l|cc||}
\hline
                  & $E_{AFDMC}$ & $E_{exp}$ \\
\hline
\hline
E($^{42}$Ca)-E($^{48}$Ca) & -55.1(8) & -54.09  \\
E($^{43}$Ca)-E($^{48}$Ca) & -46.4(8) & -46.16  \\
E($^{44}$Ca)-E($^{48}$Ca) & -34.0(8) & -35.03  \\
E($^{45}$Ca)-E($^{48}$Ca) & -27.7(8) & -27.62  \\
E($^{46}$Ca)-E($^{48}$Ca) & -17.4(8) & -17.22  \\
E($^{47}$Ca)-E($^{48}$Ca) & -10.1(8) &  -9.95  \\
\hline
\end{tabular}
\end{center}
\caption{Ground-state energy differences (in MeV) computed by AFDMC among Ca isotopes 
considered in this work.}
\label{tab:diff}
\end{table}

Most of the information needed to understand the effects of NN and TNI interaction
in the external shell can be obtained looking at energy differences between the
isotopes considered. In fact, if the intra-shell interaction has a dominant effect, the gaps
should not depend too much on the quality of the external well considered.

\begin{figure}
\begin{center}
\vspace{0.5cm}
\includegraphics[angle=0,scale=0.35]{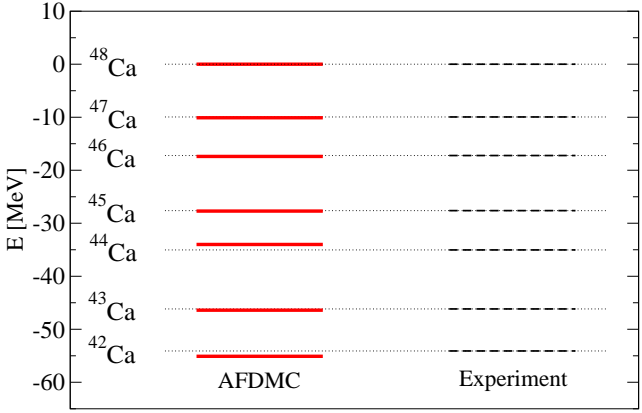}
\vspace{0.5cm}
\caption{Outline of differences between energies of the isotope series studied.
All the energies are expressed in MeV and all values are referred to the ground-state
energy of $^{48}$Ca from AFDMC and experiment respectively.}
\label{figure:diff}
\end{center}
\end{figure}

In Table \ref{tab:diff} and in Figure \ref{figure:diff} we report
the energy differences  for the isotope series considered, compared with the corresponding
differences obtained from the experimental results.
As it can be seen, in this case the agreement between computed and experimental values
is excellent. Small deviations are present only for the two cases of isotopes $^{42}$Ca 
and $^{44}$Ca.

\begin{figure}
\begin{center}
\vspace{0.9truecm}
\includegraphics[angle=0,scale=0.35]{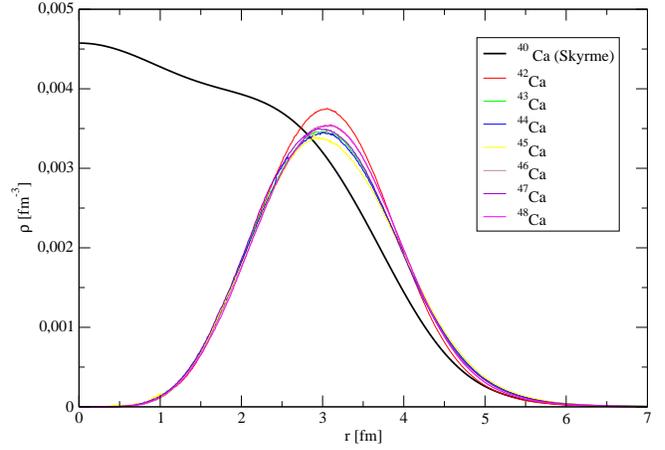}
\vspace{0.5cm}
\caption{Densities of external neutrons for all isotopes in the ground-state calculated
with AFDMC and the Skyrme density of $^{40}$Ca.}
\label{figure:dens}
\end{center}
\end{figure}

In figure \ref{figure:dens} we report the AFDMC densities normalized to unity
of the external neutrons for the isotopes considered in this work.  In the figure 
we also display the density of $^{40}$Ca calculated with the Skyrme I force. As it 
can be seen the neutron's densities are all quite similar, and very small deviations 
are present. The external neutrons are very close to the core of $^{40}$Ca 
and because of this one might expect that the interaction between the core with 
external neutrons cannot be a satisfactorily described by a one-body external potential.
However, the results on the separation energies show that this effect only contributes at the 
single particle level.

\section{Conclusions}
We used the AFDMC method to study properties of off-shell neutrons in calcium isotopes 
including a realistic two- and three-body interaction. The approximation of considering 
the core as a single particle interaction on external neutrons gives satisfactory results. In fact 
it does not affect the differences between energies of the isotopes 
considered, despite there is a systematic difference of about 6-7 MeV between our 
absolute energies with experimental values.
This fact reveals that the physics of the external neutrons is dominated by the NN and TNI interactions.
The importance functions to project the ground-state with AFDMC for the isotopes considered have 
been obtained by restricting the Hilbert space to the $1F_{7/2}$ and $1F_{5/2}$ shells. 
The quality of our results suggests that other shells, which are technically harder to include 
do not need to be considered and should introduce only very small deviations in the results.
The model we considered should be applied to investigate the microscopical structure of heavier 
neutron-rich nuclei or of excited states.

We thank S. Fantoni and K. E. Schmidt for useful discussions.
Calculations were partially performed on the BEN cluster at ECT* in Trento, under a grant
for supercomputing projects, and partially on the HPC facility "WIGLAF" of the Department of Physics, 
University of Trento.

\bibliographystyle{prsty}
\bibliography{calcio}

\end{document}